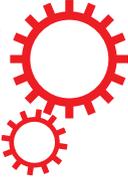

# Analytically determined topological phase diagram of the proximity-induced gap in diffusive *n*-terminal Josephson junctions

Morten Amundsen, Jabir Ali Ouassou & Jacob Linder

Multiterminal Josephson junctions have recently been proposed as a route to artificially mimic topological matter with the distinct advantage that its properties can be controlled via the superconducting phase difference, giving rise to Weyl points in 4-terminal geometries. A key goal is to accurately determine when the system makes a transition from a gapped to non-gapped state as a function of the phase differences in the system, the latter effectively playing the role of quasiparticle momenta in conventional topological matter. We here determine the proximity gap phase diagram of diffusive n-terminal Josephson junctions ($n \in \mathbb{N}$), both numerically and analytically, by identifying a class of solutions to the Usadel equation at zero energy in the full proximity effect regime. We present an analytical equation which provides the phase diagram for an arbitrary number of terminals *n*. After briefly demonstrating the validity of the analytical approach in the previously studied 2- and 3-terminal cases, we focus on the 4-terminal case and map out the regimes where the electronic excitations in the system are gapped and non-gapped, respectively, demonstrating also in this case full agreement between the analytical and numerical approach.

The interest in topological quantum phases of matter has grown steadily in recent years, and the fundamental importance of this topic in physics was recently recognized by Thouless, Haldane, and Kosterlitz being awarded the 2016 Nobel prize in physics for their contribution to this field. So far, specific material classes such as telluride-based quantum wells (HgTe, CdTe), bismuth antimony ($Bi_{1-x}Sb_x$) and bismuth selenide ($Bi_2Se_3$) have received the most attention in the pursuit of symmetry-protected topological phases and excitations[1–4]. However, it was recently proposed[5] that similar physics could be obtained using conventional superconducting materials. More specifically, by using multiterminal Josephson junctions, the authors of ref. 5 showed that it was possible to create an artificial topological material displaying Weyl singularities under appropriate conditions. In multiterminal Josephson junctions hosting well-defined Andreev bound states, the crossing of these states with the Fermi level has been shown to be analogous to Weyl points in 3D solids with the Andreev bound state taking on the role of energy bands and the superconducting phase differences corresponding to quasiparticle momenta. A considerable advantage in utilizing multiterminal Josephson junctions rather than 3D solids to study exotic phenomena such as Weyl singularities and topologically different phases is that the phase differences are much more easily controlled experimentally than the quasiparticle momenta.

In order to probe electronic excitations with topological properties, a key goal is to map out the phase diagram of the system in terms of when it is gapped or not. A gapped system here means that there are no available excitations in a finite interval around the Fermi level. The reason for why this is important is that transitions between topologically protected states can occur via gap closing, and so by identifying under which circumstances the system makes such a transition provides information about when the topological nature of the system's quantum state changes.

The arguably easiest way to probe such a phase transition is via the readily available density of states measurements, which pick up whether or not the system is gapped at a specific energy. The electronic density of states can be probed via conductance measurements, for instance in the form of tunneling between the system and a small metallic tip using so-called scanning tunneling microscopy. Recent previous works have considered the case of

Department of Physics, Norwegian University of Science and Technology, N-7491 Trondheim, Norway. Correspondence and requests for materials should be addressed to M.A. (email: morten.amundsen@ntnu.no)

3-terminal Josephson junctions, both in ballistic[6,7] and diffusive systems[8], and also the 4-terminal case in the case of chaotic cavities being connected to each other and the superconductors[9]. In particular the 4-terminal case is of interest due to the possibility of creating Weyl singularities[5].

In terms of experimental realization, metallic diffusive systems are of high relevance as the conditions for realizing such systems are far less stringent than, for instance, the discrete Andreev bound states of quantum dots. However, the proximity-gap phase diagram has not yet been studied for the 4-terminal case involving diffusive normal metals.

Motivated by this, we here determine the proximity gap phase diagram of diffusive $n$-terminal Josephson junctions ($n \in \mathbb{N}$), both numerically and analytically, by identifying a class of solutions to the Usadel equation[10] at zero energy in the full proximity effect regime. We present an analytical equation which provides the phase diagram for an arbitrary number of terminals $n$. After briefly demonstrating the validity of the analytical approach in the previously studied 2- and 3-terminal cases, we focus on the 4-terminal case and map out the regimes where the electronic excitations in the system are gapped and non-gapped, respectively, demonstrating also in this case full agreement between the analytical and numerical approach. Our results may serve as a guideline for exploring the interesting physics of multiterminal devices involving the experimentally prevalent and accessible scenario of diffusive metals connected to superconductors, which has a long history[11].

## Theory

We will use the quasiclassical theory of superconductivity which is known to yield good agreement with experimental measurements on mesoscopic superconducting devices. As only non-magnetic structures will be considered here, only singlet Cooper pairs exist and it is possible to work in Nambu-space alone due to the spin degeneracy. Using a field operator basis $\psi = (\psi_\uparrow, \psi_\downarrow^\dagger)$, the $2 \times 2$ quasiclassical Green function matrix $\underline{g}$ describing the existence of superconductivity in the system via the anomalous correlation function $f$ reads:

$$\underline{g} = \begin{pmatrix} g & f \\ \tilde{f} & -\tilde{g} \end{pmatrix} \tag{1}$$

Here, $\{g, f\}$ are complex scalars that depend on position $\mathbf{r}$ and quasiparticle energy $E$. In a bulk BCS superconductor with order parameter $\Delta = \Delta_0 e^{i\phi}$, $\underline{g}$ takes the form:

$$\underline{g}_{\text{BCS}} = \begin{pmatrix} c & s e^{i\phi} \\ -s e^{-i\phi} & -c \end{pmatrix} \tag{2}$$

where $c \equiv \cosh(\theta)$, $s \equiv \sinh(\theta)$, and $\theta = \text{atanh}[\Delta_0/(E + i\delta)]$. Here, $\delta$ accounts for inelastic scattering processes and causes a smearing of the spectral density. In writing $\underline{g}_{\text{BCS}}$, we have used that $\tilde{c} = c$ and $\tilde{s} = -s$. The above matrix may be Ricatti-parametrized[12] in the same way as one would do in the case of non-degenerate spin (see e.g. ref. 13 for a general Ricatti-parametrization in this case) with two differences: (i) we have to let $\tilde{\gamma} \to -\tilde{\gamma}$, and (ii) treat $\{\gamma, \tilde{\gamma}\}$ as scalars rather than matrices. More specifically, we write the Green function in the form

$$\underline{g} = \begin{pmatrix} N(1 - \gamma \tilde{\gamma}) & 2N\gamma \\ 2\widetilde{N}\tilde{\gamma} & -\widetilde{N}(1 - \tilde{\gamma}\gamma) \end{pmatrix} \tag{3}$$

with $N = \widetilde{N} = (1 + \gamma \tilde{\gamma})^{-1}$. The Usadel equation in the normal wires, which governs the behavior of the Green function $\underline{g}$, reads:

$$D \partial_x (\underline{g} \partial_x \underline{g}) + i[E \underline{\tau}_z, \underline{g}] = 0, \tag{4}$$

where $D$ is the diffusion coefficient, $\underline{\tau}_z$ is the third Pauli matrix, and $E$ is the quasiparticle energy. Since we are interested in mapping out the regime where the system is gapped, it suffices to consider the behavior of $\underline{g}$ at the Fermi level ($E = 0$). In this case, we have $\tilde{\gamma} = \gamma^*$, and the Ricatti-parametrized Usadel equation [obtained by inserting Eq. (3) into Eq. (4)] determining $\gamma$ takes the form

$$\partial_x^2 \gamma - \frac{2(\partial_x \gamma)^2 \gamma^*}{1 + |\gamma|^2} = 0. \tag{5}$$

This equation has the following general and exact solution if $\gamma \in \mathbb{R}$:

$$\gamma(x) = \tan(c_1 x + c_2). \tag{6}$$

Although a purely real $\gamma$ might seem like a very particular case, this scenario in fact allows us to gain important information about the proximity-gap phase diagram. To see this, consider the expression for the normalized (against its normal-state value) density of states $\mathcal{N}$ at zero energy:

$$\mathcal{N} = \frac{1 - |\gamma|^2}{1 + |\gamma|^2}. \tag{7}$$

The solution $\gamma = 0$ corresponds to the absence of superconducting correlations, i.e. completely closed gap, in which case the density of states resumes its normal-state value $\mathcal{N} = 1$. The solutions $\gamma = \pm 1$ correspond to the fully gapped case $\mathcal{N} = 0$ where no available quasiparticle excitations exist at the Fermi level. The existence of such points can now be identified analytically by determining $c_1$ and $c_2$ in Eq. (6) via the boundary conditions in the $N$-terminal system. We later proceed to do so explicitly. It is also worth noting that Eq. (5) also has a general solution when $\gamma$ is purely imaginary [$\gamma \in \mathbb{C}$, Re($\gamma$) = 0]:

$$\gamma(x) = i \tan(c_1 x + c_2). \quad (8)$$

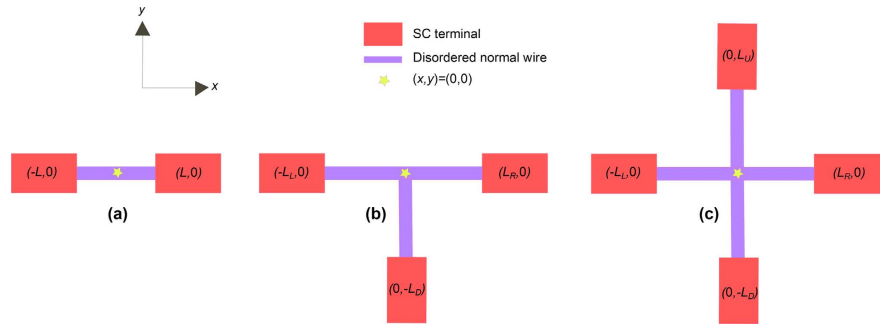

**Figure 1. Multiterminal Josephson junctions.** The density of states $\mathcal{N}$ at zero energy (Fermi level) is measured at the point indicated by a star, i.e. at the intersection of the diffusive normal wires. (**a**) 2-terminal, (**b**) 3-terminal, and (**c**) 4-terminal setups. Since the wires are assumed to be diffusive, their precise geometrical orientation does not influence the topological properties of the system. For instance, the same 3-terminal topological phase diagram would have been obtained if the leads were connected in a Y-shape rather than a T-shape: only the physical properties of the wires (*e.g.* their Thouless energies) are of consequence.

The solution Eq. (6) is of particular relevance in the case where the phase differences between the terminals is $n\pi$, with $n = 0, 1, 2, \ldots$ The reason for this is that in such a scenario, one can choose a gauge where all superconducting order parameters are purely imaginary in the reservoirs (phases $\phi_j = \pi/2$ or $3\pi/2$), which renders the BCS anomalous correlation function $f = se^{i\phi_j}$ to be purely real at zero energy since $s(E=0) = -i$. If one assumes ideal boundary conditions at the superconducting interfaces, meaning that $f$ is continuous, there are no imaginary terms in the boundary conditions or in the equation of motion for $\gamma$ itself, meaning that the solution $\gamma$ can be taken as real. From Eq. (7), it is clear that the maximum value of the Fermi-level density of states in the presence of a superconducting proximity effect is $\mathcal{N}_{\max} = 1$. We can thus conclude that the analytical approach presented above is valid whenever the superconducting phase differences between the terminals are $n\pi$.

The above class of exact solutions are useful since they are valid at specific phase differences and provide information about whether or not the DOS is gapped there. However, we have identified an additional class of exact solutions which is useful because it is valid at *any phase-differences where* $\mathcal{N} = 0$, which is precisely the regime of interest. By noting that $\mathcal{N} = 0$ only when $|\gamma| = 1$, a reasonable ansatz is:

$$\gamma = -ie^{iS(x)}, \; S(x) \in \mathbb{R}. \quad (9)$$

The prefactor $-i$ is just a convention that simplifies the boundary conditions for $S$. Insertion into Eq. (5) gives immediately

$$S(x) = ax + b \quad (10)$$

where $a$ and $b$ are real constants determined by the boundary conditions. Besides allowing us to analytically determine the region in phase-space where the system is gapped, this solution also allows us to analytically compute the topological number associated with the gapped regime defined as[14]:

$$m = \oint \nabla S(\mathbf{r}) \cdot d\mathbf{r} \quad (11)$$

where $S(\mathbf{r})$ is interpreted as the phase of the superconducting correlations at $E = 0$. There are several ways to relate the Riccati parameter $\gamma$ to the physical properties of the system. First of all, it can be related to the physically observable density of states using Eq. (7). Moreover, when the system is fully gapped so that the zero-energy density of states $\mathcal{N} = 0$, $\gamma$ is in fact just the anomalous Green function $f$, which quantifies the superconducting correlations in the system. This can be seen by comparing Eqs (1) and (3): in general, the anomalous Green function is given by $f = 2N\gamma$, but since $\gamma = -ie^{iS(\mathbf{r})}$ for a fully gapped system, we find that $N = [1 + e^{+iS(\mathbf{r})}e^{-iS(\mathbf{r})}]^{-1} = 1/2$ using the definition given above. It is assumed that the Green functions in the superconductors may be approximated by bulk expressions, and that the interfaces to the normal metals are transparent. This leads to the boundary conditions $S(\mathbf{r}_j) = \phi_j$, where $\mathbf{r}_j$ are the locations of the terminals in Fig. 1, and $\phi_j$ are the corresponding phases. This can be deduced by comparing with the anomalous Green function in a bulk superconductor, $f_{\text{BCS}} = -ie^{i\phi}$.

Although Eq. (9) is exact whenever the system is gapped ($\mathcal{N} = 0$), it cannot be used carelessly because one still has to specify for which choices of the phases $\phi_j$ it is valid. It is clearly valid when all phases are equal in the system, so that the phase-difference between all terminals is zero. As we will later show, it is also valid in large regimes of phase-space, and one needs a criterion for when Eq. (9) can be used. Such a criterion can be obtained in a convenient way by noting that as soon as $S(x)$ acquires a non-zero imaginary part, the consequence is that $\mathcal{N} \neq 0$. Identifying the condition for when a complex $S(x)$ becomes a possible solution is thus our strategy for describing analytically the topological phase diagram. By using Eq. (9) with $S(x) \in \mathbb{C}$ and writing $S(x) = S_r(x) + iS_i(x)$, Eq. (5) becomes

$$\partial_x^2 S + i(\partial_x S)^2 \left[1 - \frac{2}{1 + e^{2S_i}}\right] = 0 \qquad (12)$$

It is observed that the solution of Eq. (12) reduces to Eq. (10) in the limit $S_i(x) \to 0$. This means that by allowing a small $S_i(x)$, it is possible to map out regions where Eq. (10) is not valid and the imaginary component begins to matter. To do so, we Taylor expand the square bracket of Eq. (12), and insert the perturbation expansion

$$S(x) = S_r(x) + i(\lambda S_{i1}(x) + \lambda^2 S_{i2}(x) + \ldots) \qquad (13)$$

where $S_{i1}(x) \ll S_r(x)$ and $S_{ik+1} \ll S_{ik}$. The expansion parameter $\lambda$ is a helper variable used to collect different orders of the expansion. This gives

$$\lambda^0: \quad \partial_x^2 S_r = 0 \qquad (14)$$

$$\lambda^1: \quad \partial_x^2 S_{i1} + (\partial_x S_r)^2 S_{i1} = 0 \qquad (15)$$

and similarly for higher orders of $\lambda$. It is noticed in particular that Eq. (10) remains a solution for $S_r(x)$. The first order correction $S_{i1}(x)$ is easily solved, giving

$$S_{i1}(x) = C_1 \cos(ax) + C_2 \sin(ax) \qquad (16)$$

In an $n$-terminal Josephson junction with transparent interface between superconductors and the normal metal, it is clear that $|\gamma| = 1$ at the interface regardless of the phase. The proper boundary conditions are therefore that $S_{i1}(x_j) = 0$, with $x_j$ being the position of superconducting interface $j$. In addition, current conservation at the intersection between the arms of the multiterminal junction requires continuity of the Green function as well as the following relation between derivatives:

$$\sum_j \vec{e}_j \cdot \nabla \gamma_j = 0 \qquad (17)$$

where $\gamma_j$ is the solution of the Usadel equation in arm $j$, and $\vec{e}_j$ is a unit vector pointing towards the intersection. Using these conditions, it is possible to formulate a criterion for when the purely real solution for $S(x)$ is valid, namely: Any combination of boundary conditions for which the only solution for $S_{i1}(x)$ possible is one where $C_1 = C_2 = 0$. The curves where this is *not* satisfied may be found from the boundary conditions for an $n$-terminal Josephson junction as

$$\sum_{j=1}^{n} \frac{\psi_j}{\tan \psi_j} = 0 \qquad (18)$$

with $\psi_j$ given as

$$\psi_j = \phi_j - \langle \phi \rangle = \phi_j - \frac{1}{n}\sum_{k=1}^{n} \phi_k \qquad (19)$$

Equations (18) and (19) represent a key analytical result in this manuscript as they provide the phase diagram for the proximity-induced gap for an arbitrary number of terminals $n$. It is emphasized that the curves satisfying Eq. (18) only determine when a small imaginary contribution to $S(x)$ is possible and hence for which phases a transition between gapped and ungapped regimes in phase space occur. These curves are therefore referred to as transition curves. Higher order terms in the perturbation expansion are required in order to more accurately describe the ungapped regions. This is however not necessary when only interested in the gapped regions. It will be shown that it is possible to distinguish between the two regimes using only the first order correction.

To complement our analytical considerations, we also perform a fully numerical determination of the proximity-gap phase diagram by solving the Usadel equation numerically for any phase differences $\phi_j$ and without assuming ideal boundary conditions. In the following sections, we first provide a brief discussion of the already known 2-terminal and 3-terminal cases in order to prove the correctness of our novel analytical approach. Then, we proceed to discuss the less explored 4-terminal case in more detail.

We comment here that multiterminal geometries beyond effective 1D models can also be treated using the recently developed[15] numerical solution of the full Usadel equation in 3D, allowing for the study of non-trivial geometrical effects. Moreover, previous works have considered analytical solutions of the Usadel equation using the so-called $\theta$-parametrization in SN bilayers[16–18] and also approximate solutions in the SNS case[19–21], whereas in our work the analytical solution is *exact* for the key cases of *(i)* $\mathcal{N} = 0$ and *(ii)* for phase differences $n\pi$ between the terminals.

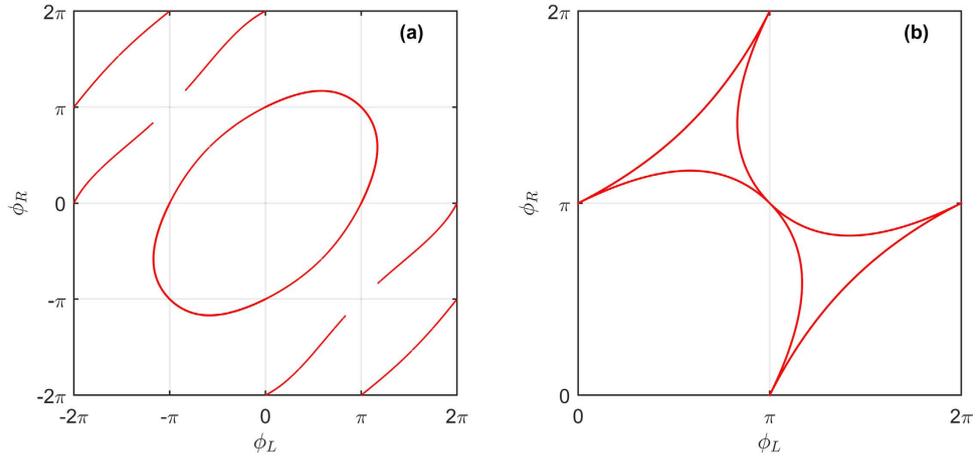

**Figure 2. Analytically calculated transition curves between gapped and ungapped regions in the 3-terminal case.** Plot of curves where the first order correction $S_{i1}(x)$ can have non-zero solutions. (**a**) Structure of the condition in the extended phase space, showing metastable solutions. (**b**) Translation of physically relevant curves into $[0, 2\pi] \times [0, 2\pi]$.

## Results: 2-terminal case

Assuming ideal boundary conditions at the superconducting interfaces $x = -L/2$ and $x = L/2$ see Fig. 1(a), real solutions of $\gamma$ must satisfy $\gamma = \tan(c_1 x + c_2)$ where:

$$\tan(-c_1 L/2 + c_2) = -ie^{i\phi_L} \text{ and } \tan(c_1 L/2 + c_2) = -ie^{i\phi_R} \quad (20)$$

This restricts the superconducting phases to be $\phi_j = \{\pi/2, 3\pi/2\}$ in order to ensure $\gamma \in \mathbb{R}$. A number of solutions can be obtained from this. If $\phi_L = \pi/2$ and $\phi_R = 3\pi/2$ or vice versa, the solution is $c_2 = 0$ which gives a DOS in the center of the wire $\mathcal{N}(x=0) = 1$. This is the expected result for a phase difference of $\pi$ between the superconducting terminals. If instead the phase difference is zero, meaning $\phi_L = \phi_R = \{\pi/2, 3\pi/2\}$, then the solution is $c_2 = \pm \pi/4$, providing $\mathcal{N}(x=0) = 0$. This is also consistent with the result that the DOS is allowed to be fully gapped when there is no phase difference. These results are in agreement with the condition given in Eq. (18), which identifies $\phi_L - \phi_R = n\pi$, $n = 1, 2, \ldots$ as the only configurations for which a non-zero density of states is possible. The phase-dependent minigap in an SNS junction was originally considered in ref. 19.

## Results: 3-terminal case

In the 3-terminal case, we consider the geometry of Fig. 1(b). The regions in phase space where $\mathcal{N}(x=0, y=0) = 0$ is mapped out using Eq. (18). Since only phase differences matter physically, we fix the phase of one superconducting terminal, $\phi_D = 0$, without loss of generality. Transition curves indicating the transition between gapped and ungapped regions are shown in Fig. 2(a) for the extended phase space $[-2\pi, 2\pi] \times [-2\pi, 2\pi]$. It can be seen that one such curve encircles the origin, with a near-elliptical shape, thereby splitting the plane into two regions. It is known that the origin resides in a gapped region, so that the outer region may be identified as ungapped. There also appears several open curves in the second and fourth quadrant. These curves are considered to be metastable solutions, corresponding to a higher phase-winding of the superconducting correlations in the normal wires, and are not investigated further. Due to the $2\pi$-periodicity of the superconducting phases, the physically relevant transition curves must be translated into $[0, 2\pi] \times [0, 2\pi]$, as shown in Fig. 2(b).

The density of states may also be computed analytically in the select points where the boundary conditions are real. Using Eq. (6), the solutions in the left, down, and right arm are written as $\gamma_L = \tan(c_1 x + c_2)$, $\gamma_R = \tan(c_3 x + c_4)$, $\gamma_D = \tan(c_5 x + c_6)$. For this particular calculation, it is necessary to set $\phi_D = \pi/2$ in order to make $\gamma_{\text{BCS},D} = -ie^{i\phi_D} = 1$ real. At the intersection point $(x, y) = (0, 0)$ continuity of the Green function and its derivative ensure continuity of the current. We assume here for simplicity equal lengths and normal-state conductances of the three normal wires, although the analytical treatment does not require this in general. In this case, we obtain the boundary conditions

$$\tan(-c_1 L_L + c_2) = -ie^{i\phi_L}, \quad \tan(-c_5 L_D + c_2) = 1,$$
$$\tan(c_3 L_R + c_2) = -ie^{i\phi_R}, \quad (1 + \tan^2 c_2)(c_1 + c_5 - c_3) = 0. \quad (21)$$

The values of $\{\phi_L, \phi_R\}$ are restricted to $\pi/2$ and $3\pi/2$ in order to ensure the validity of the solution for $\gamma$. Since $\tan c_2 \in \mathbb{R}$, the last boundary condition is equivalent to $c_1 + c_5 - c_3 = 0$. The above non-linear system of equations may be solved analytically, keeping the physically acceptable solution which gives $\mathcal{N} > 0$. For instance, for $(\phi_L, \phi_R) = (3\pi/2, 3\pi/2)$ one finds that $\tan(c_2) = -2 \pm \sqrt{3}$. The positive solution is the physically acceptable one since it provides $\mathcal{N} > 0$. The Fermi-level DOS in the center of the system $(x, y) = (0, 0)$ is given by

$$\mathcal{N}(x=0, y=0) = \frac{1 - |\tan(c_2)|^2}{1 + |\tan(c_2)|^2}, \quad (22)$$

and we find from the solution of $c_2$ that:

$$\mathcal{N}(x=0, y=0) = \begin{cases} 0, & \text{if } (\phi_L, \phi_R) = (\pi/2, \pi/2) \\ 0.866, & \text{if } (\phi_L, \phi_R) = (\pi/2, 3\pi/2) \\ 0.866, & \text{if } (\phi_L, \phi_R) = (3\pi/2, \pi/2) \\ 0.866, & \text{if } (\phi_L, \phi_R) = (3\pi/2, 3\pi/2) \end{cases} \quad (23)$$

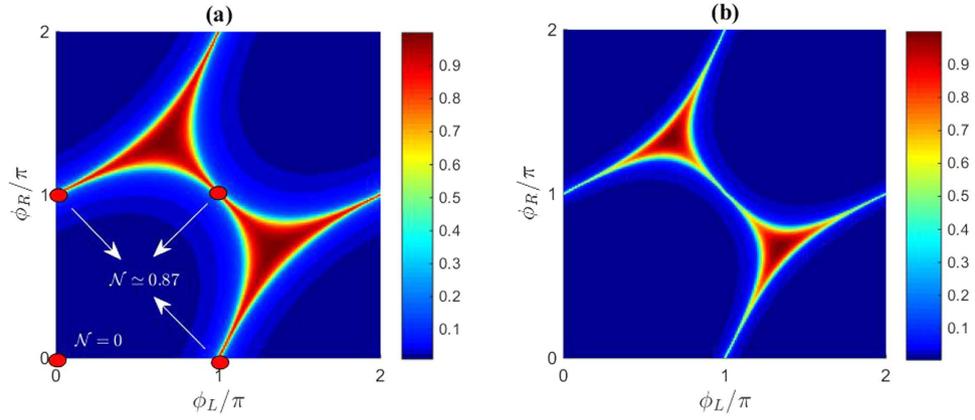

**Figure 3. Numerically calculated proximity-gap phase diagram for 3-terminal Josephson junctions.** Plot of the Fermi level density of states $\mathcal{N}$ for a 3-terminal setup as a function of the phases $\phi_L$ and $\phi_R$. For both plots, we set $L/\xi = 0.67$ and $\delta/\Delta_0 = 5 \times 10^{-3}$. The phase of the 'down' superconducting terminal has been set to $\phi_D = 0$. (**a**) Ideal boundary conditions. (**b**) Kupriyanov-Lukichev boundary conditions with finite interface resistance. We have set $\zeta_j = 2.5$, $j = \{L, R, D\}$.

These solutions may be compared with the numerical solution of the full proximity-gap phase diagram in Fig. 3(a), where it can be seen that the analytically determined transition curves of Fig. 2(b) trace out exactly the regions where the density of states is non-zero. The excellent correspondance is explained by the rapid transition between the two regimes, as shown by the numerical solution. In addition, the four red circles are gauge-equivalent to the above phase-choices (note that in the figure we have set $\phi_D = 0$). As seen, the analytical expressions match the numerical result. In order to model a more realistic setting with finite interface transparencies, we provide the phase diagram using the Kupriyanov-Lukichev boundary conditions[22] in Fig. 3(b). The interface transparency is quantified by the parameter $\zeta_j = R_{B,j}/R_{N,j}$ where $R_{Bj}$ is the barrier resistance and $R_{Nj}$ is the normal-state resistance of wire $j$. As seen, the gapped region extends compared to the fully transparent case, in agreement with ref. 8.

### Results: 4-terminal case

We now focus on the 4-terminal case and demonstrate both the robustness of the analytical approach developed above in addition to providing comprehensive numerical results. The transition surface in the, now three dimensional, extended phase space is shown in Fig. 4(a), where $\phi_U$ has been fixed to zero and metastable solutions have been removed for clarity. It can be seen to have an ellipsoidal shape, which is an expected generalization of the 3-terminal case. Figure 4(b–d) show slices of the surface after translation into the first quadrant for $\phi_D = 0, \frac{\pi}{2}$ and $\pi$, respectively. The resulting phase diagram displays a more complicated behavior than in the 3-terminal case. At $\phi_D = 0$, the phase diagram is similar to the 3-terminal case, but as $\phi_D$ is increased toward $\pi/2$ one of the gapped regions expands greatly at the expense of the other gapped regions which are separated from each other by a "barrier" of finite DOS $\mathcal{N} \neq 0$. As $\phi_D$ is further increased toward $\pi$, the phase-diagram morphs into a qualitatively different shape than at $\phi_D = 0$, and at $\phi_D = \pi$ two of the gapped regions have been almost completely expelled from the phase diagram whereas two gapped "valleys" remain, the latter again separated by a non-gapped region.

With purely real boundary conditions, and $\phi_U = \frac{\pi}{2}$, the solutions in the left, down, right, and up arm are written as $\gamma_L = \tan(c_1 x + c_2)$, $\gamma_D = \tan(c_3 x + c_4)$, $\gamma_R = \tan(c_5 x + c_6)$, $\gamma_U = \tan(c_7 x + c_8)$. As in the previous section, we assume here for simplicity equal lengths and normal-state conductances of the four normal wires. The resulting boundary conditions take the form:

$$\tan(-c_1 L_L + c_2) = -ie^{i\phi_L}, \ \tan(-c_5 L_D + c_2) = -ie^{i\phi_D}, \ \tan(c_3 L_R + c_2) = -ie^{i\phi_R},$$
$$\tan(c_7 L_U + c_2) = 1, \ (c_1 + c_5 - c_3 - c_7) = 0. \quad (24)$$

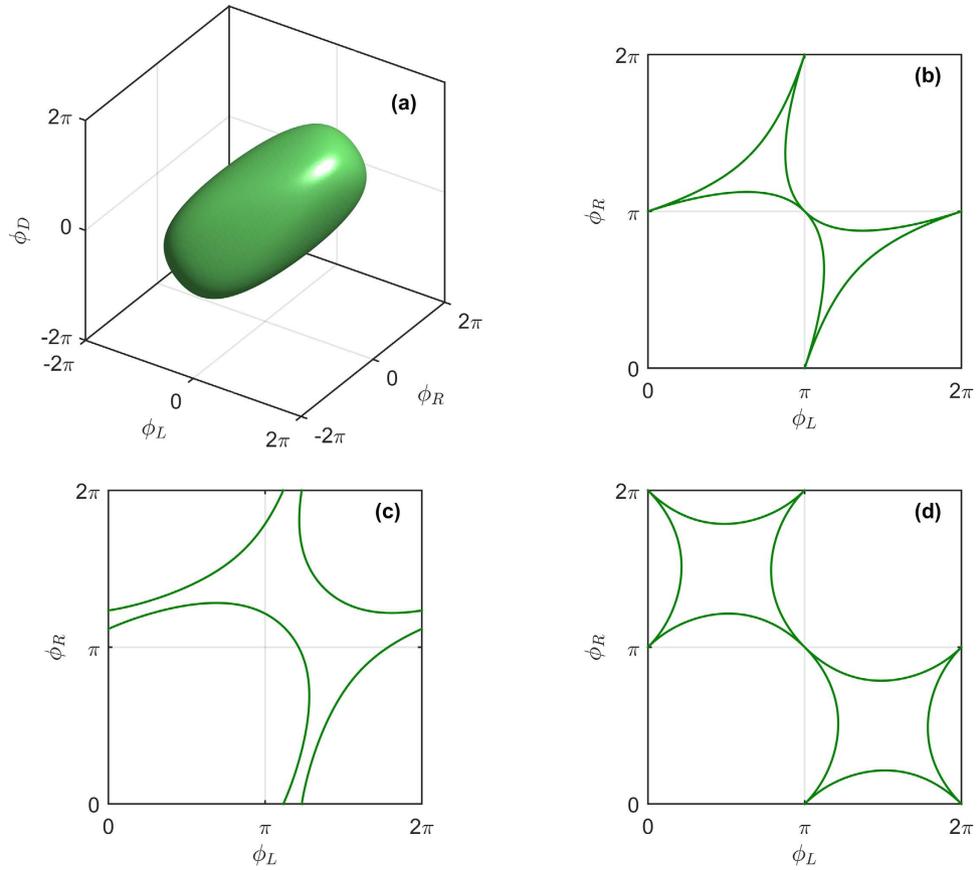

**Figure 4. Analytically calculated transition curves between gapped and ungapped regions in the 4-terminal case.** The mapping of three-dimensional phase space was performed using Eq. (18), with $\phi_U = 0$. (**a**) Transition surface in extended phase space. (**b–d**) Translation of physically relevant curves into the first quadrant for $\phi_D = 0, \frac{\pi}{2}$ and $\pi$, respectively.

|  | $(\phi_L, \phi_R) = (\pi/2, \pi/2)$ | $(\phi_L, \phi_R) = (3\pi/2, \pi/2)$ | $(\phi_L, \phi_R) = (\pi/2, 3\pi/2)$ | $(\phi_L, \phi_R) = (3\pi/2, 3\pi/2)$ |
|---|---|---|---|---|
| $\phi_D = \pi/2$ | $\mathcal{N} = 0.00$ | $\mathcal{N} = 0.71$ | $\mathcal{N} = 0.71$ | $\mathcal{N} = 1.00$ |
| $\phi_D = 3\pi/2$ | $\mathcal{N} = 0.71$ | $\mathcal{N} = 1.00$ | $\mathcal{N} = 1.00$ | $\mathcal{N} = 0.71$ |

**Table 1. Analytically obtained values of $\mathcal{N}$ at special points in phase-space.** The solution for the zero-energy DOS $\mathcal{N}$ at the intersection point of the wires $(x, y) = (0, 0)$ obtained through analytically solving the non-linear equations for $\gamma_j$ assuming transparent interfaces to the superconducting terminals (in contrast to Figs 5 and 6 where a finite interface resistance is used). We fixed $\phi_U = \pi/2$. At all points $(\phi_U, \phi_D, \phi_L, \phi_R)$ shown in the table, the analytically obtained value of $\mathcal{N}$ matches the numerically obtained solution.

This non-linear system of equations may be solved analytically. Due to the requirement that $\gamma \in \mathbb{R}$, we restrict our attention to $\{\phi_L, \phi_R, \phi_U\}$ taking the values $\pi/2$ and $3\pi/2$. We provide the solutions in Table 1 which again match the values obtained from a fully numerical solution, thus indicating the correctness of our analytical approach.

We now proceed to present numerical results for the 4-terminal case when there exists a finite interface resistance between the superconducting terminals and the normal wires, which is experimentally more realistic. We fix $\phi_U = 0$ without loss of generality and plot the evolution of the proximity-gap phase diagram, quantified via the zero-energy DOS $\mathcal{N}$ at the intersection point $(x, y) = (0, 0)$, as the remaining superconducting phases $\{\phi_D, \phi_L, \phi_R\}$ are varied in Fig. 5. Once again, the analytical transition curves correspond well with the regions where the numerically computed density of states differs from zero.

In an experimental setting, the phase-differences can be tuned by connecting the superconducting terminals and thus creating loops which a magnetic flux can pass through, the latter controlling $\phi_j$. We consider in Fig. 6 the special case where the flux penetrating all loops is the same, meaning that the phase difference between each pair of terminals is equal to $\phi$ (except between the up and left terminal, see inset of Fig. 6). We set all wire lengths $L_j = L$ and interface resistances to be equal for simplicity, and consider different sizes $L$. Regardless of $L$, the superconducting correlations vanish completely at $\phi = \pi/2$ and $\phi = \pi$, as indicated by $\mathcal{N}$ taking its normal state value ($\mathcal{N} = 1$). The gapped region at $0 < \phi < \pi/2$ for small lengths $L/\xi \ll 1$ starts to fill up with available electronic excitations as $L$ increases.

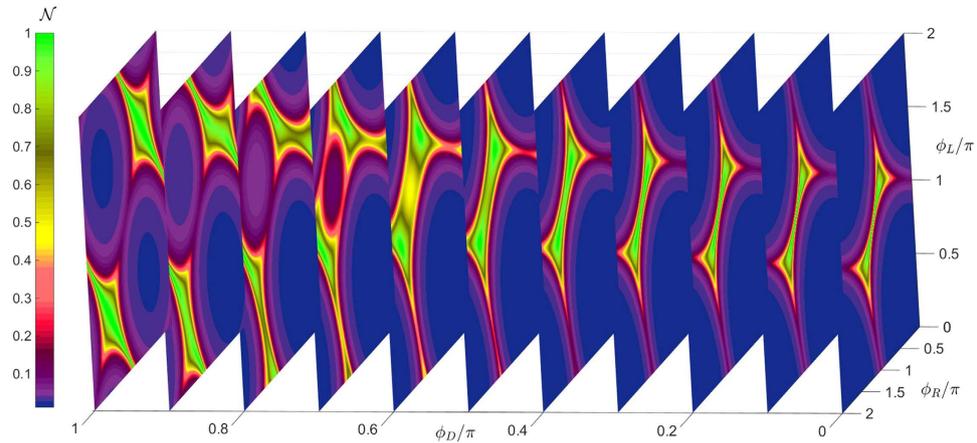

**Figure 5. Numerically calculated density of states at $E = 0$ for a 4-terminal Josephson junction for different phase-configurations.** Setting the upper superconducting phase to zero without loss of generality, $\phi_U = 0$, we plot the evolution of the proximity-gap phase diagram, quantified via the zero-energy density of states $\mathcal{N}$ at the intersection between the wires, as the phases at the other superconducting terminals are varied. We have set the wire lengths equal to $L/\xi = 0.67$ and the interface contact with the superconductors parametrized by a finite interface resistance ratio to the bulk resistance $\zeta = 2.5$. The blue regions correspond to the gapped regime where $\mathcal{N} = 0$.

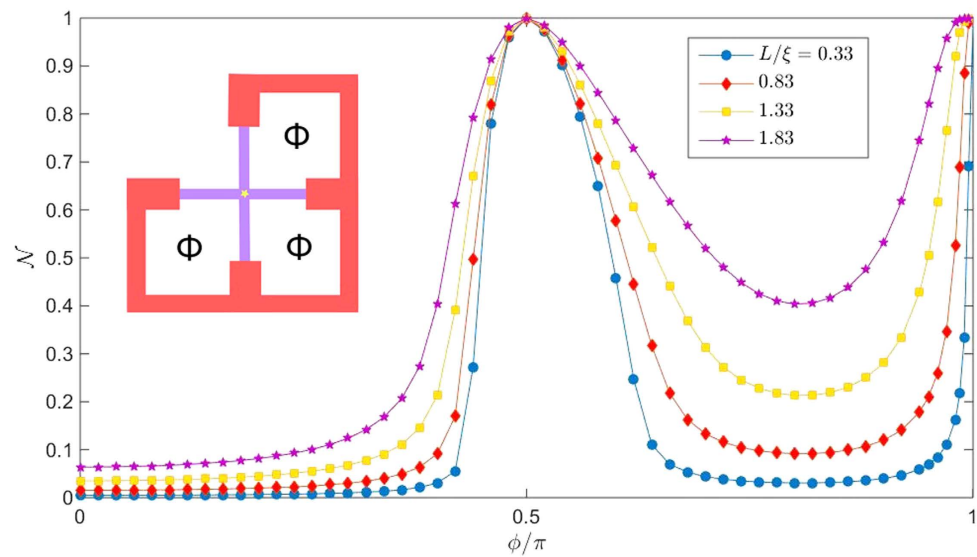

**Figure 6. Numerically calculated density of states at $E = 0$ for a 4-terminal Josephson junction for equal flux through the loops.** Plot of the Fermi level density of states $\mathcal{N}$ for a 4-terminal setup as a function of $\phi$ where $\phi_R = \phi$, $\phi_D = 2\phi$, $\phi_L = 3\phi$, which corresponds to a scenario where the same flux $\Phi$ penetrates loops that connects the superconducting terminals (see inset). We have set $\phi_U = 0$ without loss of generality, $\delta/\Delta_0 = 3 \times 10^{-3}$, and $\zeta_j = 2.5$, $j = \{L, R, D, U\}$.

## Conclusion

The main new results in this work are the class of analytical solutions of the Ricatti-parametrized Usadel equation at $E = 0$ in the full proximity effect regime, the equations (18) and (19) providing the transition between the gapped and non-gapped regimes for an arbitrary number of terminals $n$, and the specific results for the 4-terminal case. An interesting expansion of the present work would be to explore how magnetic interfaces[23–25] and spin-orbit coupling would influence the proximity-gap phase diagram and topological properties of multi-terminal Josephson junctions, as recent works have demonstrated that in particular the latter of these can induce several novel effects in both diffusive and ballistic superconducting hybrids[13,26–34].

## Acknowledgements

J.L. was supported by the Research Council of Norway, Grant No. 216700 and the "Outstanding Academic Fellows" programme at NTNU. J.L. and J.A.O. were supported by the Research Council of Norway, Grant No. 240806.

## Author Contributions

J.L. conceived the idea of the project and performed the initial analytical and numerical calculations with input from J.A.O. and M.A. The majority of the results for the analytical solution of the Ricatti-equation and belonging phase-diagram were obtained and refined by M.A. with support from J.L. and J.A.O. All authors contributed to the discussion and writing of the manuscript.

## Additional Information

**Competing financial interests:** The authors declare no competing financial interests.